\documentclass{PoS}
\usepackage{amsmath}

\title{Results from Electron-Positron Collisions at BESIII}

\ShortTitle{Results from BESIII}

\author{\speaker{Ryan E. Mitchell}\thanks{On behalf of the BESIII Collaboration.}\\
        Indiana University\\
        E-mail: \email{remitche@indiana.edu}}

\abstract{
A selection of results from electron-positron collisions at BESIII are reviewed.  The results presented here illustrate the wide range of physics topics that can be studied using the Beijing Electron Positron Collider~(BEPC).  At low collision energies, the cross section for $e^+e^-\to \pi^+\pi^-$ provides much-needed input into theoretical calculations of the anomalous magnetic moment of the muon, and the reaction $e^+e^-\to p\bar{p}$ provides access to the electromagnetic form factors of the proton.  In the charmonium region, a large sample of $\psi^\prime$ decays can be used to measure new decay modes of charmonium states.  And at higher energies, BESIII is uniquely situated to explore questions concerning the still-unexplained $XYZ$ states.
}

\FullConference{38th International Conference on High Energy Physics\\
		3-10 August 2016\\
		Chicago, USA}

\begin{document}

BESIII has collected a variety of data sets for $e^+e^-$ collisions with center-of-mass energies between 2.0 and 4.6~GeV~\cite{Briere:2016puj}.  A few of the highlights, from low to high energy, include: a scan of the region between 2.0 and 3.0~GeV; 1.3~billion $J/\psi$ decays; 450~million $\psi^\prime$ decays; 2.9~fb$^{-1}$ of data at the $\psi(3770)$ mass; about 3~fb$^{-1}$ at 4.18~GeV~(primarily for studies of the $D_s$ meson); 0.8~fb$^{-1}$ in a scan of the region between 3.85 and 4.59~GeV~(spread over 104~points); and over 4~fb$^{-1}$ collected between 3.81 and 4.60~GeV (in sets ranging from 50~pb$^{-1}$ to 1.1~fb$^{-1}$) for studies of the $XYZ$ states.  In addition to these fixed energies, one can also study $e^+e^-$ collisions at any lower center-of-mass energy using the Initial State Radiation~(ISR) technique, where photons are radiated from the primary $e^+$ or $e^-$ before the collision.  The following represents a small selection of the recent results that have been derived from these data sets.

\section{
Anomalous Magnetic Moment of the Muon}

\begin{figure}[htb]
\includegraphics*[width= 0.42\columnwidth]{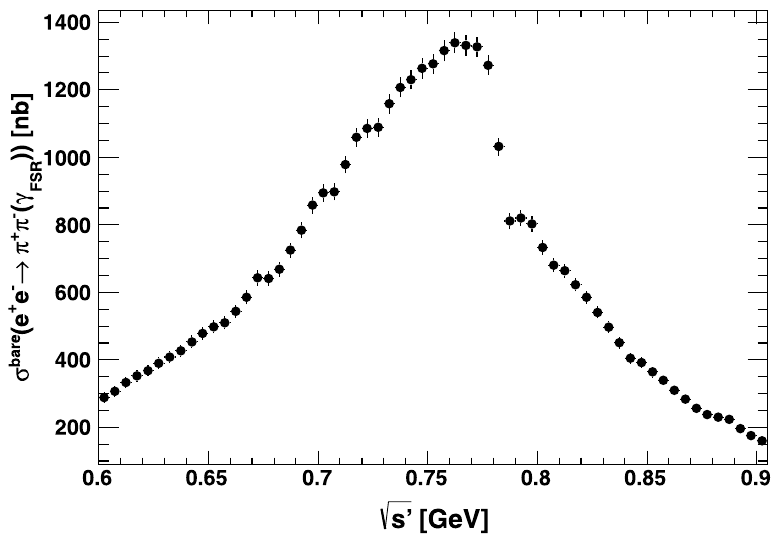}
\includegraphics*[width= 0.58\columnwidth]{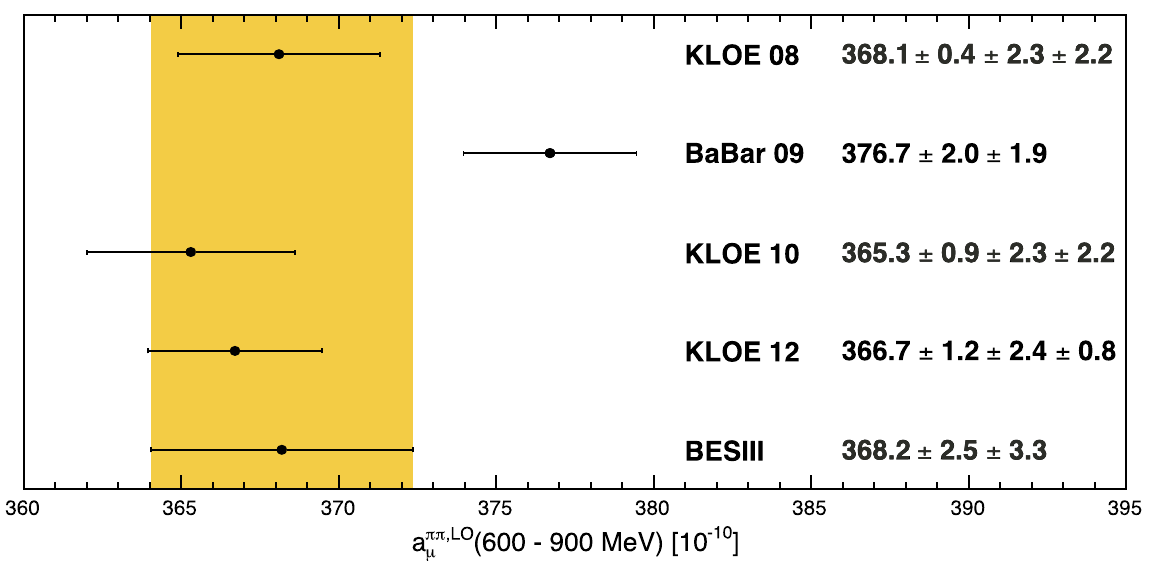}
\caption{\label{fig:1}(left)~The cross section for $e^+e^- \to \pi^+\pi^-$~\cite{Ablikim:2015orh}.  (right)~A comparison of measurements of $a_\mu^{\pi\pi\mathrm{LO}}$(600-900~MeV), where the yellow band illustrates the new BESIII measurement. }
\end{figure}

The difference between the Standard Model~(SM) and the experimental~(E821) values for the anomalous magnetic moment of the muon, $a_\mu \equiv (g_\mu-2)/2$, is currently larger than 3$\sigma$: 
\begin{align*}
a_\mu^{\mathrm{SM}} &= (11659180.2 \pm 4.9) \times 10^{−10}~\cite{Davier:2010nc}, \\
a_\mu^{\mathrm{E821}} &= (11659209.1 \pm 6.3) \times 10^{−10}~\cite{Olive:2016xmw},\\
\Delta a_\mu = a_\mu^{\mathrm{E821}} - a_\mu^{\mathrm{SM}} &= (28.9 \pm 8.0) \times 10^{−10}~~(3.6\sigma).
\end{align*}
The error in the SM calculation is dominated by the Hadronic Vacuum Polarization~(HVP) contribution, which is estimated using experimental input from $e^+e^-$ collisions to hadrons. 
The cross section for $e^+e^-$ collisions to hadrons, in turn, is dominated by the reaction $e^+e^− \to \pi^+\pi^-$ in the region of the $\rho$ meson, corresponding to collision energies between 600 and 900~MeV.
But here there are experimental differences between BaBar and KLOE, as shown in Figure~\ref{fig:1}b, on the order of several sigma.  If only the BaBar measurement is used in $a_\mu^{\mathrm{SM}}$, the difference between SM and experiment drops below 3$\sigma$.  It is thus crucial to provide more experimental input.

This discrepancy between BaBar and KLOE was addressed at BESIII using 2.9~fb$^{-1}$ of $e^+e^-$ data at a nominal center-of-mass energy of 3.773~GeV~\cite{Ablikim:2015orh}.  The ISR technique was used to measure the cross section for $e^+e^- \to \pi^+\pi^-$ in the region between 600 and 900~MeV~(Fig.~\ref{fig:1}a).
This was then integrated using a dispersion relation to obtain a new value for $a_\mu^{\pi\pi\mathrm{LO}}$(600-900~MeV)~(Fig.~\ref{fig:1}b).  The BESIII measurement, by favoring KLOE, confirms the existence of a larger than $3\sigma$ deviation of $\Delta a_\mu$ from zero.

\section{
Electromagnetic Form Factors of the Proton}

\begin{figure}[htb]
\begin{center}
\includegraphics*[width= 1.0\columnwidth]{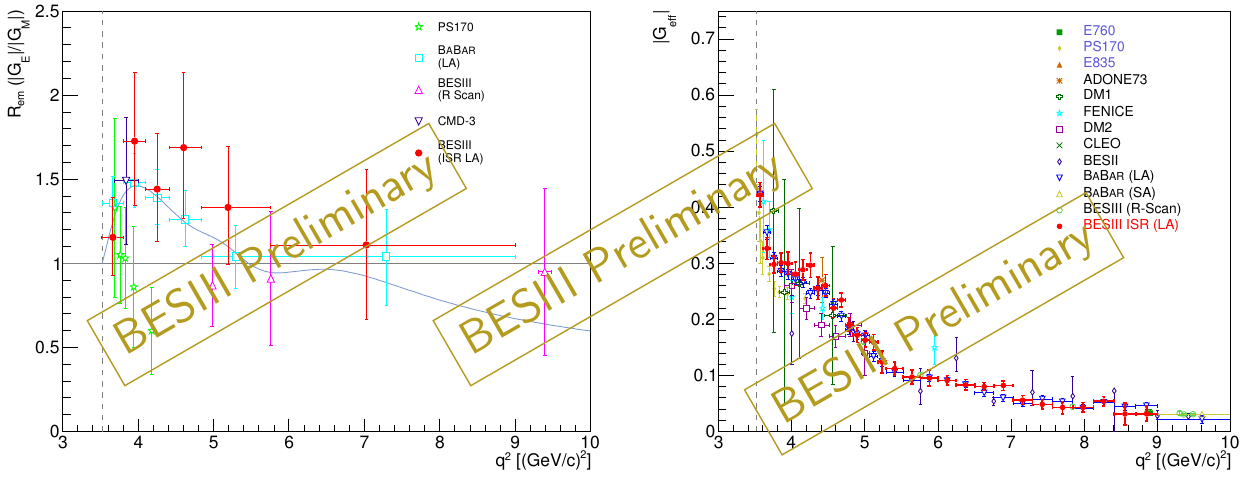}
\caption{\label{fig:2}(left)~The ratio $|G_E|/|G_M|$ in the timelike region as recently measured by BESIII~(red).  (right)~The value for $|G_{eff}|\equiv|G_E|=|G_M|$ as recently measured by BESIII~(red).  Previous measurements are shown using other colors.  See, for example, Ref.~\cite{Ablikim:2015vga}.  The line is from the Phokhara event generator.}
\end{center}
\end{figure}

The electromagnetic form factors of the proton can be measured in the spacelike region (where the momentum transfer, $q^2$, is less than zero) using elastic scattering of the electron off of the proton, $e^-p \to e^-p$.  The same form factors can also be studied in the timelike region ($q^2>0$) using the corresponding reaction $e^+e^- \to p\bar{p}$.  
BESIII has preliminary results for $e^+e^-\to p\bar{p}$ covering a wide range of collision energies, obtained using the ISR technique, starting with seven data samples at higher energies with a total integrated luminosity of 7.4~fb$^{-1}$.  The form factors $|G_E(q^2)|$ and $|G_M(q^2)|$ are measured by binning the data in $q^2$~(which in this case is equivalent to the center-of-mass energy of the collision) and fitting the distribution of the scattering angle of the proton~($\theta_p$) using:
\begin{equation*}
\frac{d\sigma_{p\bar{p}}(q^2)}{d\cos\theta_p}=\frac{2\pi\alpha^2\beta C}{4q^2}\left[|G_M(q^2)|^2(1+\cos^2\theta_p)+
\frac{4m_p^2}{q^2}|G_E(q^2)|^2\sin^2\theta_p\right],
\end{equation*}
where $\alpha$ is the fine structure constant, $\beta$ is the proton velocity, and $C$ is a Coulomb correction factor.
The results for the ratio $|G_E|/|G_M|$ are shown (in red) in Figure~\ref{fig:2}a.  Assuming $|G_E|=|G_M|\equiv|G_{eff}|$, the results for $|G_{eff}|$ are shown (in red) in Figure~\ref{fig:2}b.  

\section{
Studies of Charmonium}

One of the most interesting problems in charmonium physics is the unexpected difference between decays of the $J/\psi$ and $\psi^\prime$ to light hadrons.  This was first noticed in the $\rho\pi$ system, where the $\psi^\prime$ decay to $\rho\pi$ is suppressed relative to expectations based on the corresponding $J/\psi$ decay~\cite{Brambilla:2010cs}.  But related phenomena have since been seen in many more channels.  One striking example is in radiative decays to the $\eta$ and $\eta^\prime$.  
While the ratio of branching fractions, $B(\gamma\eta)/B(\gamma\eta^\prime)$, is $21.9\pm0.9$\% for the $J/\psi$, it is only $1.1\pm0.4$\% for the $\psi^\prime$~\cite{Olive:2016xmw}.

\begin{figure}[htb]
\begin{center}
\includegraphics*[width= 0.6\columnwidth]{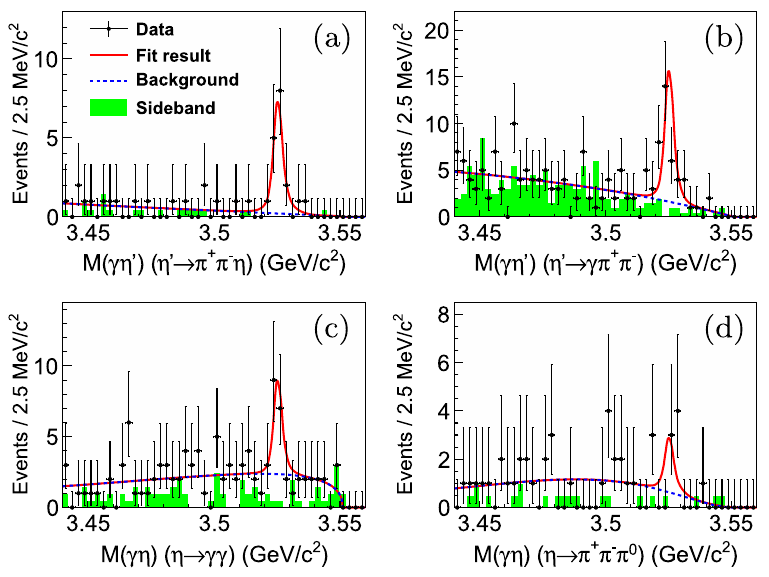}
\caption{Measurement of $h_c\to\gamma\eta^{(\prime)}$ for the decays (a)~$\eta^\prime\to\pi^+\pi^-\eta$, (b)~$\eta^\prime\to\gamma\pi^+\pi^-$, 
(c)~$\eta\to\gamma\gamma$, and (d)~$\eta\to\pi^+\pi^-\pi^0$~\cite{Ablikim:2016uoc}.
\label{fig:3}}
\end{center}
\end{figure}

Using a sample of 450~million $\psi^\prime$ decays, BESIII has been able to make a measurement of the same ratio, $B(\gamma\eta)/B(\gamma\eta^\prime)$, for $h_c$ decays~\cite{Ablikim:2016uoc}.  
The processes $\psi^\prime\to\pi^0h_c$ with $h_c\to\gamma\eta^{(\prime)}$ were reconstructed using two decay modes of the $\eta^\prime$ ($\pi^+\pi^-\eta$~(Fig.~\ref{fig:3}a) and $\gamma\pi^+\pi^-$~(Fig.~\ref{fig:3}b)) and two decay modes of the $\eta$ ($\gamma\gamma$~(Fig.~\ref{fig:3}c) and $\pi^+\pi^-\pi^0$~(Fig.~\ref{fig:3}d)).  
Simultaneous fits were performed for the two $\eta$ and $\eta^\prime$ channels and the ratio was measured to be $B(\gamma\eta)/B(\gamma\eta^\prime) = 30.7 \pm 11.3 \pm 8.7$\%, showing that the $h_c$ decays behave more like the decays of the $J/\psi$ than the $\psi^\prime$.

\section{
Results on the ``Y'' States}

\begin{figure}[htb]
\begin{center}
\includegraphics*[width= 0.9\columnwidth]{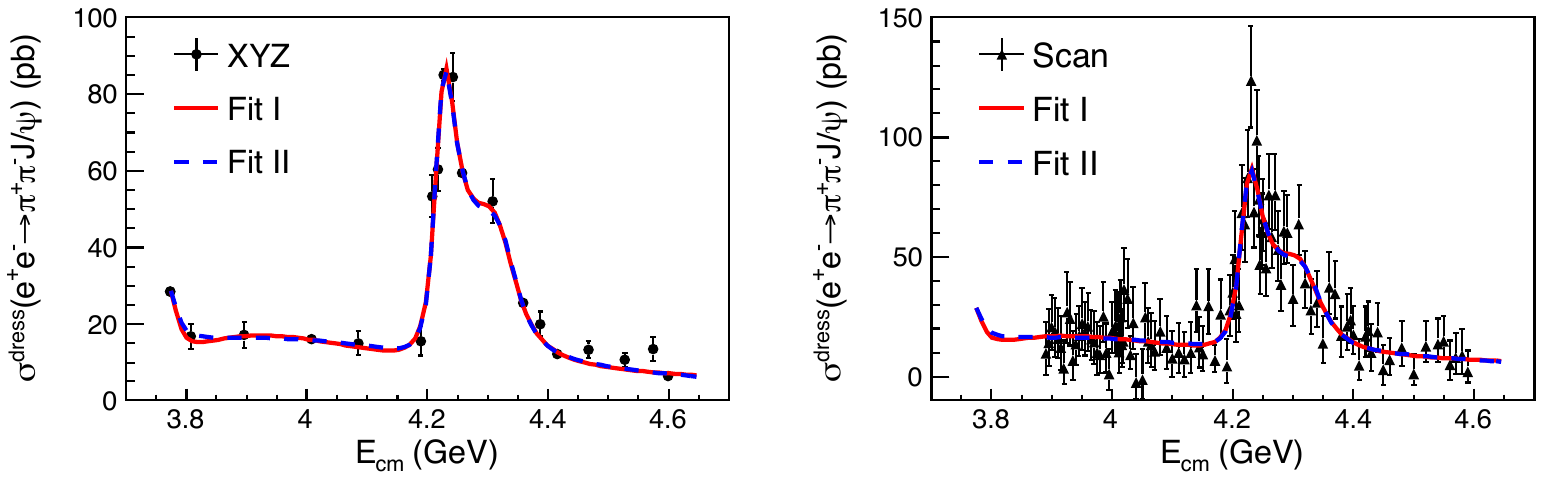}
\includegraphics*[width= 0.45\columnwidth]{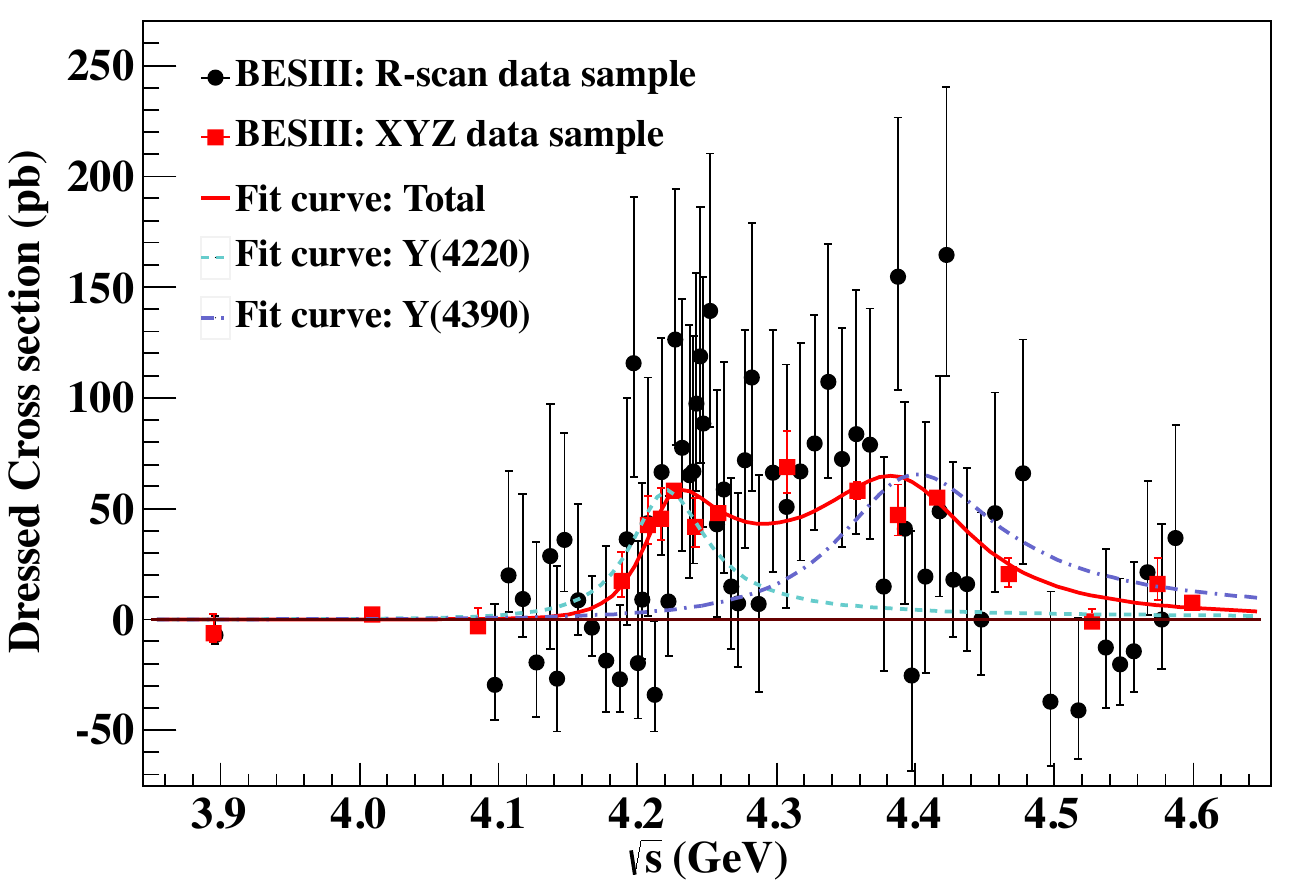}
\includegraphics*[width= 0.45\columnwidth]{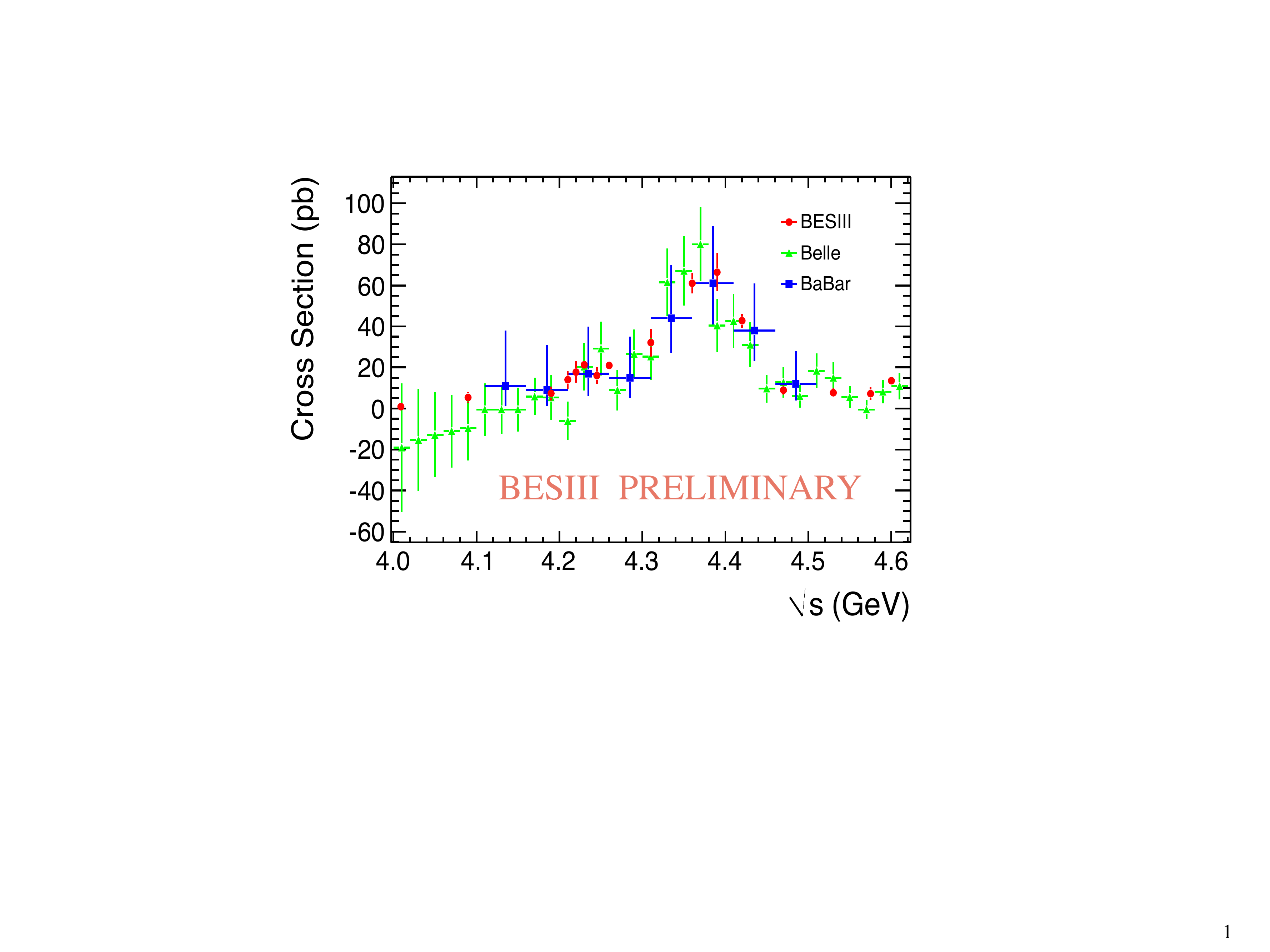}
\caption{Measurements of $e^+e^-$ cross sections to exclusive final states including charmonium at BESIII.  (top~left)~The $e^+e^-\to\pi^+\pi^-J/\psi$ cross section measured using high-statistics data points~\cite{Ablikim:2016qzw};  (top~right)~the same using a larger number of low-statistics data points;  (bottom~left)~the $e^+e^-\to\pi^+\pi^-h_c$ cross section~\cite{BESIII:2016adj};  (bottom~right)~preliminary results for the $e^+e^-\to\pi^+\pi^-\psi^\prime$ cross section.
\label{fig:4}}
\end{center}
\end{figure}

Above the threshold to produce open charm, $e^+e^-$ cross sections to final states including charmonium show many puzzling features~\cite{Lebed:2016hpi}.  The first to be discovered, and the best known of these features, is the $Y(4260)$, which was originally seen as a peak in the $e^+e^-\to\pi^+\pi^-J/\psi$ cross section at a mass of around 4.26~GeV.  New results from BESIII, however, show that the $Y(4260)$ is not a simple peak~\cite{Ablikim:2016qzw}.  The BESIII measurement of the $e^+e^-\to\pi^+\pi^-J/\psi$ cross section, measured using both a small number of high-statistics data points and a large number of low-statistics data points, is shown in the top plots of Figure~\ref{fig:4}.  The peak that was formerly known as the $Y(4260)$, can, in fact, be better described as a combination of two peaks, 
one with a mass of $4222.0 \pm 3.1 \pm 1.4$~MeV/$c^2$ and width of $44.1 \pm 4.3 \pm 2.0$~MeV and 
one with a mass of $4320.0 \pm 10.4 \pm 7.0$~MeV/$c^2$ and width of $101.4^{+25.3}_{-19.7} \pm 10.2$~MeV.

Similarly, the $e^+e^-\to \pi^+\pi^-h_c$ cross section is much more complex than previously suspected.  The latest measurement from BESIII is also shown in Figure~\ref{fig:4}~\cite{BESIII:2016adj}.  It can also be described as two peaks, one with a mass of 
$4218.4 \pm 4.0 \pm 0.9$~MeV/$c^2$ and width of $66.0 \pm 9.0 \pm 0.4$~MeV
and one with a mass of 
$4391.6 \pm 6.3 \pm 1.0$~MeV/$c^2$ and width of $139.5 \pm 16.1 \pm 0.6$~MeV.
The parameters of the lighter peak agree with the parameters of the lighter peak in $\pi^+\pi^-J/\psi$.  Whether or not they originate from the same resonance is a question that requires more investigation.

Finally, BESIII has a new preliminary result on the shape of the $e^+e^- \to \pi^+\pi^- \psi^\prime$ cross section, also shown in Figure~\ref{fig:4}.  The new measurements are in agreement with those from Belle and BaBar, which were used to determine the parameters of the $Y(4360)$~\cite{Lebed:2016hpi}.

\section{ 
Results on the ``X'' States}

\begin{figure}[htb]
\begin{center}
\includegraphics*[width= 0.40\columnwidth]{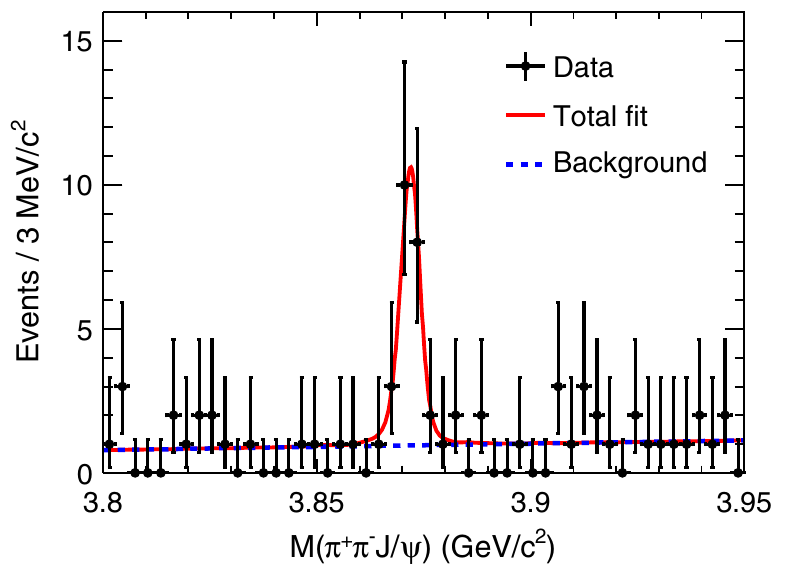}
\includegraphics*[width= 0.42\columnwidth]{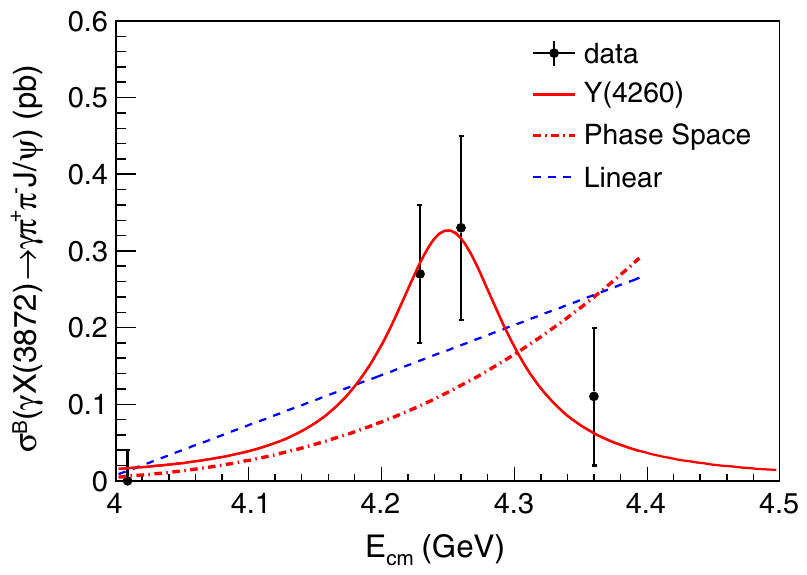}
\end{center}
\caption{
(left)~Observation of the $X(3872)$ in the process $e^+e^-\to \gamma X(3872)$ with $X(3872)\to\pi^+\pi^-J/\psi$~\cite{Ablikim:2013dyn}.  (right)~The $e^+e^-\to\gamma X(3872)$ cross section as a function of center-of-mass energy~\cite{Ablikim:2013dyn}.
\label{fig:5}}
\end{figure}

Besides the $Y(4260)$, another well-known mystery in the charmonium spectrum is the nature of the $X(3872)$~\cite{Lebed:2016hpi}.  BESIII has recently shown that there may be a connection between them.  Using high-statistics data points at 4.01, 4.23, 4.26, and 4.36~GeV, BESIII observed the process $e^+e^-\to \gamma X(3872)$, where the $X(3872)$ decays to $\pi^+\pi^-J/\psi$~(Fig.~\ref{fig:5}a)~\cite{Ablikim:2013dyn}.
The cross section as a function of center-of-mass energy~(Fig~\ref{fig:5}b) shows a shape that is consistent with a peak between 4.2 and 4.3~GeV, which may be consistent with the structure seen in other channels.  More data is needed to resolve this issue, but finding a connection between the $X(3872)$ and the ``Y'' states seen in $e^+e^-$ cross sections is a promising lead.

Similarly, BESIII searched for the $X(4140)$ (also known as the $Y(4140)$) in the analogous process $e^+e^-\to \gamma X(4140)$ with $X(4140)\to\phi J/\psi$~\cite{Ablikim:2014atq}.  Upper limits were set on the product $\sigma(e^+e^-\to \gamma X(4140))\times B(X(4140)\to\phi J/\psi)$ that are of the same order of magnitude as the measurements of $\sigma(e^+e^-\to \gamma X(3872))\times B(X(3872)\to\pi^+\pi^- J/\psi)$.

\end{document}